# The effect of δ-hydride on the micromechanical deformation of Zircaloy-4 studied by in situ high angular resolution electron backscatter diffraction

Siyang Wang[a,*], Szilvia Kalácska[b], Xavier Maeder[b], Johann Michler[b], Finn Giuliani[a], T. Ben Britton[a]

[a] Department of Materials, Royal School of Mines, Imperial College London, London, SW7 2AZ, UK

[b] EMPA, Swiss Federal Laboratories for Materials Science and Technology, Laboratory for Mechanics of Materials and Nanostructures, Feuerwerkerstrasse 39, 3602, Thun, Switzerland

*Corresponding author: siyang.wang15@imperial.ac.uk

## Abstract

Zircaloy-4 is used extensively as nuclear fuel cladding materials and hydride embrittlement is a major failure mechanism. To explore the effect of δ-hydride on plastic deformation and performance of Zircaloy-4, *in situ* high angular resolution electron backscatter diffraction (HR-EBSD) was used to quantify stress and geometrically necessary dislocation (GND) density during bending tests of hydride-free and hydride-containing single crystal Zircaloy-4 microcantilevers. Results suggest that while the stress applied was accommodated by plastic slip in the hydride-free cantilever, the hydride-containing cantilever showed precipitation-induced GND pile-up at hydride-matrix interface pre-deformation, and considerable locally-increasing GND density under tensile stress upon plastic deformation.

Zirconium alloys are used as fuel cladding materials in water-based nuclear reactors. The plastic deformation modes in hexagonal close packed (HCP) α-Zr include plastic slip on the prismatic, basal and pyramidal crystallographic planes as well as twinning [1]. The ratio of critical resolved shear stresses (CRSS) for <a> prismatic to <a> basal to <c+a> first order pyramidal slip in α-Zr is 1:1.3:3.5 at room temperature [2]. Under service conditions, Zircaloy claddings can pick up hydrogen from the coolant water. Diffusion of hydrogen atoms in Zr is fast at elevated temperature [3], resulting in a wide distribution of hydrogen solutes. Precipitation of zirconium hydride occurs when the hydrogen content reaches the solubility limit which varies strongly with temperature [4]. Depending upon the cooling rate and hydrogen content, four different zirconium hydride phases, including trigonal ζ phase ($ZrH_{0.25-0.5}$), face-centred tetragonal (FCT) γ phase (c/a>1, $ZrH_{1.0}$), face-centred cubic (FCC) δ phase ($ZrH_{1.5-1.66}$), and FCT ε phase (c/a<1, $ZrH_{1.75-2}$), can form [5–9]. Among these, δ-hydride is the most commonly observed phase which normally takes the form of lens-shaped packets habiting on the $\{10\bar{1}7\}$ planes of the parent α-Zr matrix [10,11] and is detrimental to the ductility and toughness of the cladding material [12–15].

Thermal expansion of the interior $UO_2$ pellets subjects cladding tubes to hoop stress during operation [16]. Therefore, hydrides oriented radially are more vulnerable to mode 1 fracture than those oriented circumferentially. As hydrides form along specific texture components in Zr alloys [17–19], the texture of the fuel claddings is often manipulated such that hydrides form circumferentially during reactor operation [20,21]. However, hydride reorientation caused by stresses during thermal cycling makes the hydrides radially-oriented reducing cladding toughness [18,20–22]. Together with the propensity of hydrogen diffusion towards the crack tip at service temperature this can ultimately lead to cladding failure through delayed hydride cracking (DHC) [23–26]. However, although experimental evidences have shown the embrittling effect of hydrides on Zr alloys, detailed knowledge of the microstructural failure mechanisms are unclear; in particular, in how the hydrides affect the accommodation of local plastic deformation which is fundamental to the understanding of the initial stage of DHC.

Compared to macroscale mechanical tests on polycrystalline bulk specimens, microscale testing enables the extraction of single crystal mechanical behaviours [1,27]. Better knowledge of the local stress states in single crystal tests can reduce the ambiguity in understanding deformation mechanisms. Weekes *et al.* [28] carried out compression tests on single crystal Zircaloy-4 micropillars containing hydride packets ~45° to the loading axis and observed that the hydrides, as located on the plane of maximum shear, can accommodate a noticeable fraction of the plastic strain imposed. They also found that slip bands initiated in the matrix can be terminated by the hydrides, implying that the hydrides are relatively strong out-of-plane.

Recently, cross correlation based, high angular resolution electron backscatter diffraction (HR-EBSD), has been used as an advanced technique for measuring variations in elastic strain (and stress), lattice rotations and geometrically necessary dislocation (GND) densities [29]. *In situ* HR-EBSD has also been used during micromechanical deformation of GaAs [30,31], commercially pure (CP) Ti [32], W [33–35], FeMnSi-based alloys [36] and Ti6242 [37]. In these experiments, EBSD scanning of the micropillar or microcantilever side surface was carried out at stages during interrupted mechanical tests and the data collected was run through post-mortem offline analytical tools to extract quantitative information characterising deformation features. We investigate here the effect of δ-hydride packet on the local strain accommodation and deformation mechanisms of Zircaloy-4, using bending tests conducted on hydride-free and hydride-containing microcantilevers.

Commercial Zircaloy-4 (Zr-1.5%Sn-0.2%Fe-0.1%Cr in weight [38]) was supplied as rolled and recrystallised plate consisting of equiaxed α-Zr grains (grain size = ~11 μm). A cuboid of about 10 mm × 10 mm × 1.5 mm was heat treated at 800 °C for 336 h in Ar atmosphere, which resulted in the formation of large 'blocky alpha' grains with an average size of over 200 μm [39]. Electrochemical hydrogen charging of the sample was performed in a solution of 65 °C, 1.5 wt% sulfuric acid using a current density of 2 kA/m$^2$, followed by an annealing at 400 °C for 5 h to homogenise the hydrogen distribution and a controlled slow furnace cooling of 1 °C/min to promote formation of δ-hydrides [40–42]. The microstructure of the sample after hydrogen charging is shown in Figure 1(a), together with the hydride nucleation sites.

A sharp 90° edge (top and cross section) was created using mechanical polishing, finishing with colloidal silica. In the near-edge region, conventional electron backscatter diffraction (EBSD) was used to map the grain orientations on the top surface, in a FEI Quanta 650 scanning electron microscope (SEM) with a Bruker eFlash$^{HD}$ EBSD camera, using a beam acceleration voltage of 20 kV. Microcantilever fabrication was conducted using Ga focussed ion beam (FIB) milling in the grain of interest at the edge of the sample, in a FEI Helios Nanolab 600 FIB-SEM. FIB with acceleration voltage of 30 kV and beam currents of 9 nA, 3 nA and 1 nA were successively used. Within one grain two microcantilevers of the same size (6 μm × 6 μm x 22 μm) were fabricated, with the c-axis of the HCP matrix nearly parallel to the cantilever long axis. Microcantilever 1, as shown in Figure 1(b), is a hydride-free single crystal Zircaloy-4 cantilever. Microcantilever 2, as shown in Figure 1(c), contains a thin intragranular hydride packet of around 50 nm in thickness lying vertically near the cantilever fixed end. In order to obtain high quality EBSD patterns during subsequent *in situ* characterisation, the side surfaces of the microcantilevers were polished with 30 kV, 270 pA FIB in a Tescan Lyra 3 FEG FIB-SEM immediately before the micromechanical testing.

A schematic diagram of the experimental setup is given in Figure 1(d). Microcantilever bending was performed with a displacement-controlled (10 nm/s during cantilever loading and unloading) Alemnis nanoindenter in the Tescan Lyra 3 SEM, using a conospherical indenter tip with a tip radius of 1 μm. The microcantilever was aligned with the loading axis parallel to the SEM stage and with the cantilever side surface inclined at 70° for *in situ* EBSD analysis.

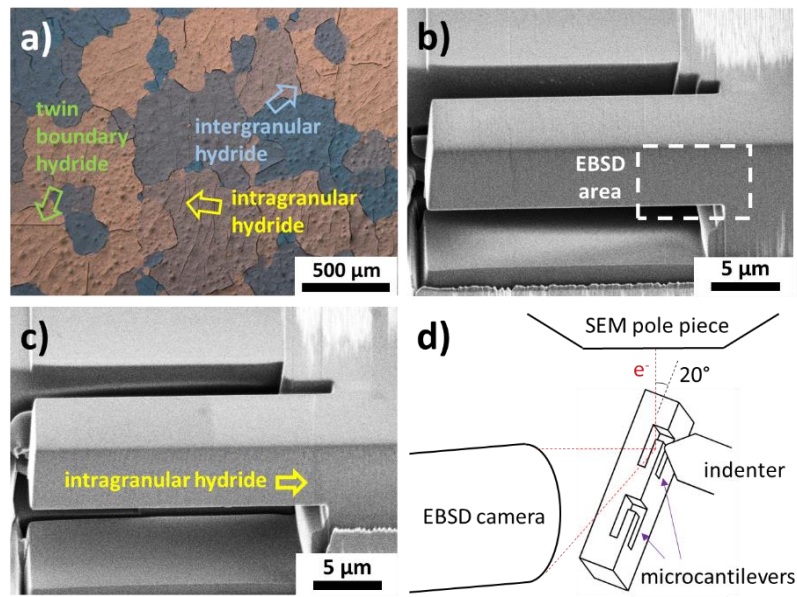

Figure 1 (a) Polarised light optical micrograph of 'blocky-alpha' large grain Zircaloy-4 with intragranular, intergranular and twin boundary hydrides. SEM images of (b) microcantilever 1, a hydride-free single crystal Zircaloy-4 and (c) microcantilever 2 in the same grain as microcantilever 1 but contains a vertical intragranular hydride near the fixed end. (d) Schematic diagram of the experimental setup for *in situ* HR-EBSD microcantilever bending test.

For each test, five EBSD maps of the microcantilever side surface were acquired with a spatial step size of 100 nm. The first and the last maps were acquired before and after (in the fully unloaded state) the tests respectively, and three others were obtained while the displacement was held constant at certain stages during the test processes. EBSD patterns were collected with an EDAX Digiview camera using a beam acceleration voltage of 20 kV and were binned to a resolution of 442 × 442 pixels. CrossCourt software v.4.255 was used for the HR-EBSD evaluation. Patterns with low image quality were discarded from the analysis. Reference patterns were selected from the support region of each microcantilever, which was likely to be at zero stress. The elastic constants of α-Zr were taken from [43] for the elastic stress calculation. Details of the principle of HR-EBSD can be found in the literature [44–46].

Figure 2 shows the load-displacement curves recorded during the bending tests of the two microcantilevers, and the variation in normal stress along the cantilever long axis ($\sigma_{xx}$) distribution maps at stages during the tests (as labelled on the load-displacement curves correspondingly) derived through HR-EBSD analysis. The other stress components are not displayed since they are small in magnitude compared to $\sigma_{xx}$ and show negligible difference between the two cantilevers. For the as-received cantilever, $\sigma_{xx}$ values prior to the test (map 1, Figure 2) are around 0 and show negligible spatial variation across the microcantilever side surface. Upon elastic deformation, tensile (positive) and compressive (negative) $\sigma_{xx}$ are present in the top and bottom regions of the surface examined (map 2), and the $\sigma_{xx}$ at the outer fibres ($\sigma_{xx,max,HR-EBSD}$) at the fixed end is around 800 MPa, in good agreement with $\sigma_{xx,max}$ extracted from the load-displacement response (830±30 MPa - see Appendix for elastic bending stress calculations). The load value upon yielding for the as-received cantilever (as denoted by the cross on the curve) is approximately 2.12 mN. The maximum bending (normal) stress at the fixed end $\sigma_{xx,max,yield}$ is therefore ~1.17 GPa (see Appendix). According to the crystal orientation of the microcantilevers (Figure 2(a)), for a uniaxial stress applied along the x direction, the highest Schmid factors for <a> prismatic, <a> basal, <a> pyramidal, <c+a> first order and second order pyramidal slip systems are 0.001, 0.012, 0.019, 0.424 and 0.466 respectively. Moreover, as <c+a> slip takes place more readily on first than second order pyramidal planes in low c/a ratio HCP metals such as Ti and Zr [47], <c+a> first order pyramidal slip is the easiest slip mode in the present loading configuration for single crystal (as-received) Zr and is therefore responsible for the initiation of plastic deformation. The CRSS for <c+a> first order pyramidal slip was hence extracted through multiplying the $\sigma_{xx,max,yield}$ of the as-received cantilever by the Schmid factor for the slip system and was found to be ~496 MPa. This CRSS value is in good agreement with the size-independent CRSS for <c+a> first order pyramidal slip in CP Zr (532±58 MPa) [2].

Further deformation into the plastic regime has led to progressively higher stress levels on the cantilever side surface (maps 3 and 4, Figure 2), with $\sigma_{xx}$ reaching up to ±2.4 GPa at an indenter displacement of ~1.7 μm. A V-shaped pattern is present in the compressive stress region close to the cantilever support, as highlighted with dashed lines in maps 3 and 4. The alignment between the shape of the pattern and the first order pyramidal planes as shown in Figure 2(a) suggests that local stress relief has taken place, likely due to the occurrence of

<c+a> first order pyramidal slip. When the cantilever was fully unloaded, a general decrease of $\sigma_{xx}$ can be observed (map 5).

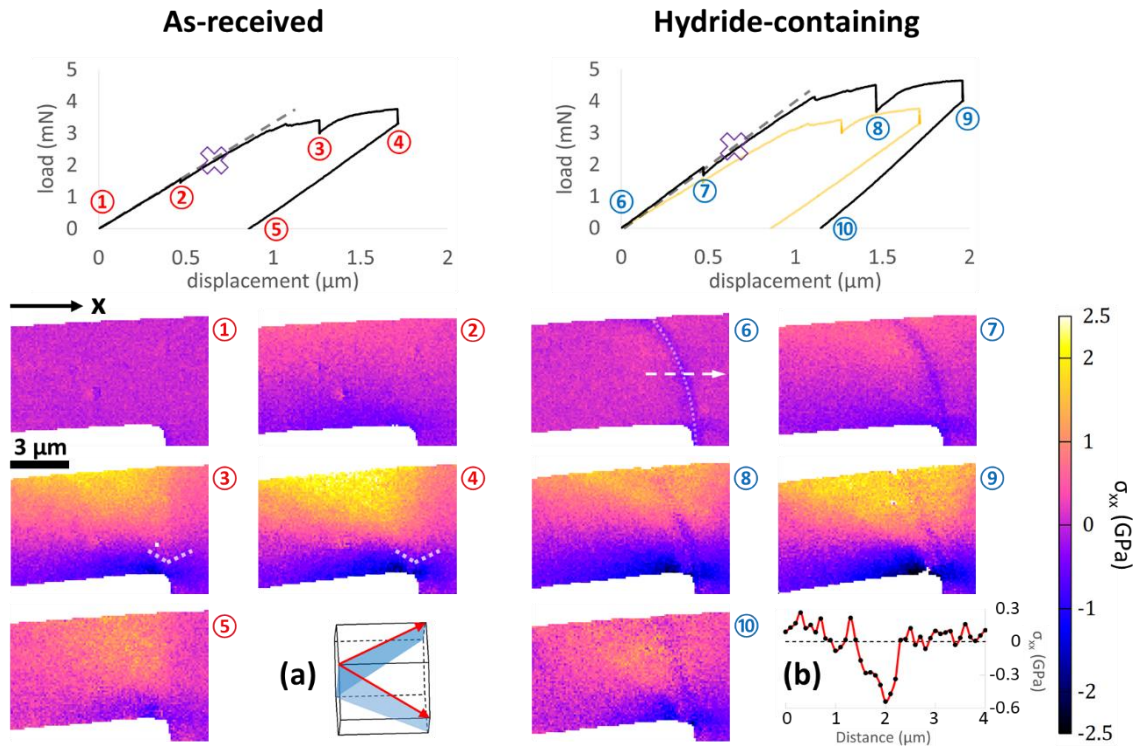

Figure 2  Load-displacement curves for bending tests of the as-received and hydride-containing microcantilevers with crosses denoting yield points (the yellow-coloured curve is that for the as-received cantilever in order for comparison), and normal stress along the cantilever long axis ($\sigma_{xx}$) distribution maps at various stages during the tests derived through HR-EBSD analysis. (a) unit cell represetation of the crystal orientation of the microcantilevers, (b) $\sigma_{xx}$ variation along the arrow on map 6.

For the hydride-containing cantilever, a compressive $\sigma_{xx}$ field is observed around the hydride packet (position of the hydride is labelled with a dotted line in Figure 2, map 6) prior to deformation. The $\sigma_{xx}$ along the arrow in map 6, as plotted in Figure 2(b), shows that the compressive stress field has a width of ~900 nm perpendicular to the hydride and a peak $\sigma_{xx}$ of ~0.5 GPa. This is likely produced upon hydride formation as a result of the misfit between the hydride and the matrix, of which the dilatational volumetric strain is ~17.2% [48]. The precipitation-induced $\sigma_{xx}$ value obtained here agrees with that extracted from finite element calculations (i.e. between 407.0 and 719.2 MPa for hydride/metal with yield stresses ranging between 200 and 500 MPa and a model hydride aspect ratio of 0.1) [49]. During elastic deformation, the precipitation-induced compressive stress in the region above the neutral axis was partly relieved (map 7). Upon plastic deformation, superposition of the precipitation-induced compressive stress and the applied tensile stress above the neutral axis resulted in the changing to a tensile field around the hydride (map 8 and 9). Below the neutral axis, the field remained compressive and increased in magnitude. After unloading (map 10), the stress value decreased in general, however, the stress around the hydride packet above the neutral axis switched to be compressive again. Throughout the deformation process, stress patterns further from the hydride are similar to those observed on the as-received cantilever.

The evolution of GND density during the deformation processes of the two microcantilevers is plotted in Figure 3. For the as-received cantilever, GND density is negligible pre-deformation and upon elastic deformation (map 1 and 2). A systematic increase in the GND density occurred upon plastic deformation (map 3) and an increase with deformation further into the plastic regime (map 4) can be seen. The GND distribution is more localised in the compressive stress region than in the tensile stress region, and local GND pile-ups along the first order pyramidal planes can be observed near the cantilever support. The GND density decreased in general after the cantilever was unloaded (map 5).

On the hydride-containing cantilever, GNDs are observed to pile up along the hydride-matrix interface before the test, particularly evident at the top and bottom of the surface examined (map 6, Figure 3, highlighted with arrows), which are likely produced upon hydride precipitation to accommodate the misfit strain at the hydride-matrix interface [41,50]. These GNDs then increased in density subtly upon elastic deformation (map 7) and extended along the hydride-matrix interface in the tensile stress region when the cantilever was deformed plastically (map 8 and 9). In the compressive stress region, however, no significant GND pile-up along the phase boundary can be seen. The density of GNDs, including those piled up at the matrix-hydride interface, decreased after unloading (map 10). Generally, the GND evolution in those regions further from the hydride is similar to that for the as-received cantilever (maps 1-5).

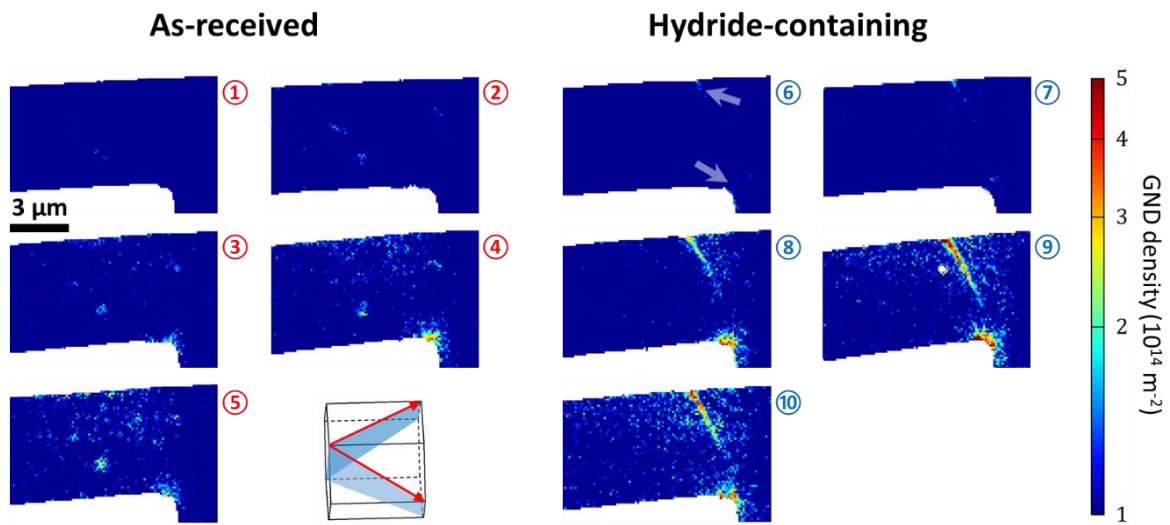

Figure 3 GND density evolution during the microcantilever bending tests, with an insert of the unit cell represetation of the crystal orientation of the microcantilevers.

In summary, bending tests were carried out on two single crystal Zircaloy-4 microcantilevers while one of them contained a hydride packet sitting vertically at the cantilever fixed end. The two cantilevers had identical crystal orientation with c-axis nearly parallel to the cantilever long axis. Under the applied stress, localised slip associated with the <c+a> first order pyramidal planes accommodated plastic deformation of the as-received cantilever. Stress and GND density evolution during the deformation process of the hydride-containing cantilever is found to be different to that of the as-received cantilever, especially in the regions around the hydride packet. A compressive stress field of ~0.5 GPa in peak magnitude and ~900 nm in width was observed in the direction perpendicular to the hydride plane, likely due to the dilatational misfit upon hydride formation. During plastic deformation, the hydride, as located perpendicular to the applied tensile stress, can lead to considerable pile-up of GNDs at the hydride-matrix interface. This may further result in both local mode 1 fracture events and hydrogen diffusion towards the dislocated area, triggering the nucleation and growth of more hydrides when the material is employed at high temperature [19]. These observations of the micromechanical mechanisms supported the understanding of the initial stage of DHC where the local plasticity is modulated by the hydride, and explained why, when and how chemical potential wells are built around the hydride for further hydride precipitation, deformation and cracking.

## Acknowledgements

TBB and SW acknowledge support from HexMat (EPSRC EP/K034332/1) and MIDAS (EPSRC EP/SO1720X). TBB thanks the Royal Academy of Engineering for funding his Research Fellowship. Some EBSD in this work was performed on the FEI Quanta SEM which was supported by the Shell AIMS UTC and is housed in the Harvey Flower EM suite at Imperial College London. The authors acknowledge Dr Johannes Ast for assistance with the *in situ* tests.

## Appendix

For free-end cantilevers with square cross-section, the maximum bending (normal) stress $\sigma_{xx,max}$ in the elastic regime is:

$$\sigma_{xx,max} = \frac{6Fl}{a^3}$$

where *F* is the load applied, *l* is the moment arm and *a* is the side length of the cantilever cross-section.

*l* value for each microcantilever was measured using post-deformation SEM images.